\newcommand{\be}[0]{\begin{equation}}
  \newcommand{\ee}[0]{\end{equation}}
\newcommand{\ba}[0]{\begin{eqnarray}}
  \newcommand{\ea}[0]{\end{eqnarray}}

\newcommand{\MSb}{{\rm \overline{MS}}}

\newcommand{\alphazero}{\overline{\alpha}_0}
\newcommand{\refeq}[1]{Eq.~(\ref{eq:#1})}

\newcommand{\sigmahat}{\hat{\sigma}}

\newcommand{\PMSa}{$\mathrm{PMS}_1$}
\newcommand{\PMSb}{$\mathrm{PMS}_2$}

\newcommand{\mean}[1]{\langle #1 \rangle}

\newcommand{\tauc}{\tau_c}
\newcommand{\tauP}{\tau_P}
\newcommand{\BP}{B_P}
\newcommand{\CP}{C_P}
\newcommand{\rhoE}{\rho_E}
\newcommand{\rhoEE}{\rho_E^{\mathrm{(ESch)}}}
\newcommand{\ykt}{y_{k_t}}
\newcommand{\yfJ}{y_{fJ}}

\newcommand{\GeV}{\mathrm{GeV}}

\documentclass[]{JHEP3}
\usepackage{epsfig} 

\title{Scale Optimisation and Event Shapes in Deep-Inelastic Scattering}
\author{M. J. Dinsdale\\ Institute for Particle Physics Phenomenology, University of Durham,
South Road, Durham, DH1 3LE, UK, \\
Institut f\"ur Physik, Johannes-Gutenberg Universit\"at,
Staudingerweg 7, D-55099 Mainz, Germany \thanks{Current address.} \\
email: \email{dinsdale@thep.physik.uni-mainz.de}}


\preprint{IPPP/05/40 DCPT/05/80 MZ-TH/05-29}

\abstract{We study the effect of optimising the renormalization and factorisation scales
on perturbative calculations of event shape means defined in the Breit frame
of ep DIS.  Unlike in the case of $e^+e^-$ event shape means, this has only a small
effect on the NLO QCD predictions and a large power
correction is still required to fit the data.  However, if
separate renormalization scales are introduced for the quark- and gluon-initiated
sub-processes the optimisation has a much larger effect and greatly reduces the size
of the required power corrections.  Unfortunately, there are then problems fitting
the low energy data for some observables.}

\begin{document}

\section{Introduction}

Event shapes are among the most powerful observables used for testing QCD \cite{Dasgupta:2003iq}, providing,
for example, some of the best measurements of $\alpha_s$ to date \cite{Bethke:2004uy}.  However these measurements
are complicated by the fact that event shapes
are unusually sensitive to soft, long-distance effects which are expected to be
beyond the reach of perturbation theory.  These effects give contributions to the mean values of the shape
variables, so-called power corrections, that only fall off like the first power of the centre of mass energy $Q$,
and remain sizable even at relatively high energies.
Indeed state-of-the-art next-to-leading order (NLO) $\MSb-$scheme perturbative
calculations give a totally unsatisfactory description of the data even for $Q$ as high as $M_Z$
without adding an additional power-suppressed contribution.

Unfortunately there is as yet no rigorous framework
for computing QCD scattering amplitudes non-perturbatively, so a phenomenological model is
required to handle the soft effects.  This can then be combined with NLO perturbative
calculations to give a successful theoretical description of the data.  

There are different approaches to modelling the power corrections.  One possibility is to use
Monte Carlo event generators to calculate the difference between ``parton-level'' and
``hadron-level'' distributions of event shapes.  More recently an analytic ansatz has
been proposed, relating the power corrections for many event shape means to a single
non-perturbative parameter, $\alphazero(\mu_I)$ \cite{Dokshitzer:1995zt,Dokshitzer:1997ew}.
This can be interpreted as the mean value of a
hypothetical universal, infrared-finite coupling $\alpha_{\mathrm{IR}}(\mu)$ over the range $0<\mu<\mu_I$.  Typically
$\mu_I$ is taken to be 2GeV, and the $\alphazero$ values required to fit the data are scattered
around 0.5, with a spread of around $20\%$ (see e.g.~the review Ref.~\cite{Dasgupta:2003iq} and references therein).
Impressively, this holds for event shape means and distributions measured  both in $e^+e^-$ annihilation and deep-inelastic
scattering (DIS).

However, in a recent analysis of $e^+e^-$ event shape means \cite{r12}, the DELPHI collaboration found that when
using an ``optimised'' choice of renormalization scale the NLO perturbative QCD calculation
could describe the data very well without any non-perturbative corrections.  
The idea behind an ``optimised'' scale is to use some sort of theoretically motivated criterion
for choosing an observable-dependent renormalization scale $\mu$ (and in general scheme) based on features of the
perturbative calculations for that observable.  The hope is that this minimises the effect of higher-order
corrections better than the standard procedure of setting the $\MSb$-scheme renormalization scheme to $Q$.
The particular scale choice used in the DELPHI
analysis was originally proposed by Grunberg; it is the consequence of applying
the Method of Effective Charges (ECH) \cite{Grunberg:1982fw} at NLO.
Several different theoretical arguments have been proposed to motivate this scale choice (going under the
names of
the Fastest Apparent Convergence criterion, Renormalization-Scheme Invariant Perturbation Theory \cite{r10},
Complete Renormalization Group Improvement \cite{Maxwell:2000pi} and Lambda-based Perturbation Theory
\cite{Dinsdale:2004yi}).

This surprising result motivates investigating the application of ``optimised'' scales to other observables which usually require large
power corrections to fit the data.  In Ref.~\cite{NLLECH}, the Method of Effective charges was applied
to the distributions of thrust and heavy-jet mass in $e^+e^-$ annihilation, including a resummation
of large logs to next-to-leading accuracy.  The optimisation did
indeed reduce the need for power corrections somewhat, although surprisingly the resummation caused problems
in the 2-jet region.

Another class of related observables, which lack the complication of large logarithms, are the event shape
means defined in the Breit frame of $ep$ DIS.  In this case, however,
the appearance of the proton in the initial state leads to a dependence of the event shape means
on the proton parton distribution functions (PDFs) and the accompanying factorisation scale, $M$.  Ideally 
we would like to use some optimisation criterion to choose all the unphysical parameters, but none
of the above-mentioned theoretical arguments simultaneously select unique values for both $\mu$ and $M$.
However, there is an alternative criterion for choosing the renormalization
scale, the Principle of Minimal Sensitivity (PMS) \cite{Stevenson:1981vj}, that does generalise very easily to the factorisation case
\cite{Politzer:1981vc,Stevenson:1986cu,Aurenche:1986ff,Aurenche:1987fs} (for recent applications to hadron-hadron interactions
see Refs.~\cite{Chyla:2003tq,Srbek:2005dy}).
Moreover, the PMS tends to agree very well with the ECH as used in the DELPHI analysis when the latter
can be applied.  Therefore, the purpose of the present paper is to apply the PMS to the choice of renormalization
and factorisation scales in DIS event shape means, and compare NLO perturbation theory to data from the H1 collaboration \cite{Herameans}
to see how the required power corrections are affected.

The plan of this paper is as follows.  Section \ref{se:Observables} contains definitions of the observables.
In Section \ref{se:scaledep} we review the way that dependence on $\mu$ and $M$ arises in NLO DIS calculations,
and define the PMS approximations we are going to study.
Then, in Section \ref{se:analysis} we describe the methods used in the calculation.  Section \ref{se:CaseStudy}
runs through the analysis in detail for one specific observable, the current jet thrust $\tauc$.
Results for all the observables discussed in Ref.~\cite{Herameans} are presented in Section \ref{se:Results}.
Section \ref{se:Conclusions} contains our conclusions.

\section{Definition of the Observables}
\label{se:Observables}

This section briefly summarises the observables which will be considerd in the paper.
For discussions of cuts and more details about the definition of 
each observable, see Ref.~\cite{Herameans}.

In all the following definitions, the momentum of the $i$th particle is $p_i$ and
Breit frame quantities are indicated by an asterisk.  The current hemisphere (CH) is defined
as that for which $z^*<0$, and the remnant hemisphere (RH) is that for which $z^*>0$.

First, we consider five observables defined using particles in the CH only.  To ensure infrared safety the total
energy in the CH (evaluated in the Breit frame) must exceed some threshold, taken to be $Q/10$, for an event
to contribute to one of these variables.  The $z^*$-axis thrust is
\be
\tauc = 1 - \frac{\sum_{i}{|\vec{p}^*_i.\vec{n}|}}{\sum_{i}{|\vec{p_i}^*|}}
\ee
where $\vec{n}$ is a unit vector in the $z^*$ direction.  Similarly, the $z^*$-axis jet broadening is
\be
\BP = \frac{\sum_{i}{|\vec{p}^*_i\;{\mathrm x}\;\vec{n}|}}{\sum_{i}{|\vec{p_i}^*|}}.
\ee
The thrust-axis thrust is
\be
\tauP = 1 - {\rm max}_{\vec{n}} \frac{\sum_{i}{|\vec{p}^*_i.\vec{n}|}}{\sum_{i}{|\vec{p_i}^*|}}
\ee
where $\vec{n}$ is now the unit vector with respect to which the maximisation is performed (the value
which leads to the maximum defines the thrust axis).  The C-parameter is defined by making use
of the tensor
\be
\Theta^*_{jk} = \frac{\sum_{i}{p^*_{i,j} p^*_{i,k} / |\vec{p_i}^*| }}{\sum_{i}{|\vec{p_i}^*|}}
\ee
with eigenvalues $\lambda_1$, $\lambda_2$ and $\lambda_3$:
\be
\CP = 3(\lambda_1 \lambda_2 + \lambda_2 \lambda_3 + \lambda_3 \lambda_1).
\ee
Lastly we have the jet mass
\be
\rhoE = \frac{(\sum_{i}{p})^2}{4 \sum_{i}{E_i^*} }.
\ee
Because the jet mass involves the difference between energies and $3$-momenta it is especially
sensitive to hadron mass effects which can lead to additional power corrections beyond the
scope of the model of Ref.~\cite{Dokshitzer:1995zt,Dokshitzer:1997ew}.  These can be removed
by defining a related observable in the so-called ``E-scheme'' \cite{Salam:2001bd}
\be
\rhoEE = \frac{\left[ \sum_{i}{(E_i^*, E_i^* \frac{\vec{p}_i^*}{|\vec{p}_i^*|})}\right] ^2}{4 \sum_{i}{E_i^*} }.
\ee
This is just $\rhoE$ with each particle replaced by a massless particle of the same energy.
Since our NLO perturbative calculations are carried out with the approximation of massless quarks, the perturbative
predictions for $\rhoEE$ will be identical to those for $\rhoE$; only the non-perturbative
contributions will differ.
For the purposes of this paper, values have been obtained for $\rhoEE$ by
applying a correction factor from PYTHIA 6.2 \cite{Pythia} to the $\rhoE$ values given in Ref.~\cite{Herameans}.

The remaining two variables make use of both the CH and the RH.  They are defined based on two jet clustering algorithms,
the factorisable Jade algorithm and the modified $k_T$ algorithm.  These clustering algorithms allow one
to gather the final-state momenta into a certain number of current jets, plus the remnant jet, according to a resolution
parameter $y$ which determines how aggressively momenta are combined.  This is done by examining the momenta repeatedly,
combining those that are nearest according to a distance measure specific to the clustering algorithm, until the nearest
pair have a separation $y_{ij}>y$.  Event shapes can be defined from these algorithms by using the value of $y$
at which the event goes from being a 2+1 jet event (two current jets, one remnant jet) to a 1+1 jet event.

In the DIS case the distance measures need to be generalised from the $e^+e^-$ case to include a distance between a jet
and the remnant.  For the factorisable Jade algorithm
{\setlength\arraycolsep{1pt}
\begin{eqnarray}
y_{ij} & = & \frac{2 E_i^* E_j^* (1 - \cos \theta_{ij}^*)}{Q^2} \\
y_{ir} & = & \frac{2 E_i^* x_B E_p^* (1 - \cos \theta_{ip}^*)}{Q^2}
\end{eqnarray}}%
where $i,j$ indicate jets, $r$ is the remnant and $p$ is the proton.  In the same notation the $k_t$ measures are
{\setlength\arraycolsep{1pt}
\begin{eqnarray}
y_{ij} & = & \frac{2 {\mathrm {min}}({E_i^*}^2, {E_j^*}^2) (1 - \cos \theta_{ij}^*)}{Q^2} \\
y_{ir} & = &\frac{2 {E_i^*}^2 (1 - \cos \theta_{ip}^*)}{Q^2}.
\end{eqnarray}}%
The related event shapes will be denoted by $\yfJ$ and $\ykt$ respectively.

\section{Applying The Principle of Minimal Sensitivity to DIS}
\label{se:scaledep}

All perturbative predictions in QCD are plagued with a dependence on the arbitrary renormalization scheme (RS) and
scale ($\mu$), together referred to as the renormalization prescription (RP) \cite{Fischer:1997bs}.
For a process with hadrons in the initial state, predictions depend also on the factorisation scheme (FS) and scale ($M$),
which we jointly refer to as the FP.  This occurs despite the fact that the exact predictions of the
theory are completely independent of these choices, because truncating the perturbation series
prevents cancellations between different orders of perturbation theory which are
necessary if the dependencies are to vanish.

Because the renormalization and factorisation prescriptions are in principle completely arbitrary,
they may be varied at will.  In the general case this would allow one to obtain any desired perturbative prediction for a given
observable.  Therefore, some method for selecting renormalization and factorisation prescriptions
is required that will give ``reasonable'' quality perturbative predictions.  In practice one normally uses the
$\MSb$ RS and FS, and sets $\mu=M=Q$.  
In this paper we are going to consider instead picking the scales using
the PMS \cite{Stevenson:1981vj}.  

%The Principle of Minimal Sensitivity is an idea of very broad applicability.  Indeed, in the paper where Stevenson
%first gave it this name and applied it to the problem of scale-dependence in QCD \cite{Stevenson:1981vj}, he noted it had already been
%in use for several years in various areas of physics.  Since then it has been put to a vast range of disparate uses.

The PMS simply states that if we are given an approximation which depends on some parameter arising only as part of the approximation
procedure (so that the exact result is independent of the parameter), the approximation is most believable where it is least
sensitive to this ``unphysical'' parameter.  This gives a simple way to choose an ``optimal'' value for the parameter (in the ideal case; in general
there may be multiple ``PMS points'' and it may be necessary to use some other criterion for choosing between them).

This principle is not the sort of thing one could imagine proving mathematically; indeed it is easy to invent approximations where
the PMS fails totally.  The sense in which the PMS gives an ``optimal'' approximation is certainly not to be understood in the mathematical
sense of minimising the difference between the approximate and exact values.  Nonetheless the PMS can be very successful
when applied to real world approximations.

In perturbative QCD one is in exactly the situation where the PMS is supposed to apply.  The unphysical parameters label the renormalization
and factorisation scales and schemes; the full sum of the perturbative series for some observable quantity is independent of these
parameters but the truncated, partial sum (which is all we have to work with in practice) is not.

The simplest application of the PMS to QCD involves the case of an observable without hadrons in the initial state (and without identified hadrons
in the final state).  The only unphysical
parameters are then those that label the renormalization scale and scheme.  At NLO it suffices to just consider the renormalization scale
$\mu$, as a change in scheme can be absorbed into a rescaling of $\mu$.  
For an observable involving hadrons in the initial state, we also have to take into account its unphysical dependence on the parameters that label
the factorisation scale ($M$) and scheme.  In contrast with the renormalization case, the factorisation scheme dependence cannot be absorbed into the scale
even at NLO.  However, it is difficult to formulate a PMS condition for the factorisation scheme in $x$-space \cite{Chyla:1988at},
so here we will simply neglect the scheme dependence and work in the $\MSb$ factorisation scheme at all times.
Hopefully this partial optimisation still represents an improvement over the physical scale approach.
With this simplification, the relevant parameters are $\mu$ and $M$.

The mean of some DIS observable $y$ depending on the final state $X$ can be expressed as
\be
\label{eq:meany}
\mean{y}=\frac{\int{y(X) d\sigma(ep\to X,Q) }}{\int{d\sigma(ep\to X,Q) }}
\ee
where $d\sigma(ep\to X,Q)$ is the infinitesimal cross-section for the process $ep\to X$, and $Q$ is the virtuality of the exchanged photon.
To compute such a cross-section in perturbation theory we need to factorise the process into a soft part, described by the proton PDFs $f_a(\xi,M)$,
and a hard part, described by a partonic cross-section $\sigmahat(e\;a \rightarrow X, Q, M)$
\be
\label{eq:PDFconvolution}
\frac{d\sigma(ep\to X, Q)}{dX} = \sum_a {\int{d\xi f_a(\xi, M) \frac{d\sigmahat(e\;a \rightarrow X, Q, M)}{dX}}}.
\ee
As indicated, $f$ and $\sigmahat$ depend
on the unphysical parameter $M$ even before they are approximated perturbatively; this
dependence cancels between them after the integral over $\xi$ and the sum over $a$ are performed.
The dependence of $f_a(\xi, M)$ on $M$ is given by the DGLAP evolution equations
\be
\frac{df_a(\xi,M)}{d\ln M} = \frac{\alpha_s(M)}{\pi} \int_\xi^1{ \frac{dz}{z} \sum_b { P_{a \leftarrow b}(z) f_b(z,M)}}.
\ee
The $P_{a \leftarrow b}(z)$ are the {\it splitting functions} (for a parton of type $b$ splitting to give a parton of type $a$ with
momentum fraction $z$).  They are uniquely defined once the factorisation scheme is specified.

To arrive at a NLO approximation, we substitute into \refeq{PDFconvolution} PDFs evolving according to the
NLO splitting functions, and the partonic cross-section expanded to NLO.  This approximation not only
prevents the exact compensation of $M$ dependence between the PDFs and the partonic cross-section,
but also introduces a dependence on the RP used for the expansion.  

In summary, at NLO, our approximation for $\mean{y}$ will have an unphysical dependence on both $M$ and $\mu$ and we can
look for a PMS point by requiring
\be
\label{eq:PMS}
\left. \frac{\partial \mean{y}_{\mathrm{NLO}}}{\partial\mu} \right|_{\mu_{\rm PMS}}=\left. \frac{\partial\mean{y}_{\mathrm{NLO}}}{\partial M} \right|_{M_{\rm PMS}}=0.
\ee
In general, the stationary point will be a saddle-point in the $(\mu,M)$ plane.  

This is the most straightforward way to apply the PMS to DIS, but the multiplicity of initial states compared
to the $e^+e^-$ case allows for some more involved possibilities.  In particular, it is interesting to consider the possibility
of using different renormalization scales in the various partonic channels.  This is legitimate as the cross-section
for each partonic sub-process is separately RP independent, as can be
seen by noting that every term in the sum over $a$ in \refeq{PDFconvolution} is independent of $\mu$ (and all
other RS parameters).
However, it must be borne in mind that these
cross-sections do depend on the {\it factorization} prescription - e.g.~the dependence on $M$ only disappears
after the sum over $a$ is performed.
Therefore, only this sum is actually physically observable.  
Now, the usual advice is to avoid adding together observables prior to optimisation (this makes sense as the PMS point could otherwise
arise as a consequence of cancellations between the two observables, which is presumably not a good indication of reliability),
but these components are not observable, so it might not be sensible to apply this rule here.  

The possibility of using different scales for the $q\gamma^*$ and $g\gamma^*$ sub-processes was also suggested
in Ref.~\cite{Ingelman:1994nz},
where the $2+1$-jet cross-section in DIS was studied using a number of ``optimisation'' methods.  There the contributions
to the jet cross-section for each sub-process were thought of as being
in principle observable (assuming one could distinguish quark and gluon jets), because of the
characteristically different final states: at Born level, the quark-initiated ``QCD compton'' process leads
to a $qg$ final state whereas the gluon-initiated ``boson-gluon fusion'' process leads to a $q\overline{q}$ final
state.  However, this
appears to contradict the fact that the cross-sections for these processes depend on the FP.  The problem
arises from considering the final state of the hard process in isolation, when the separation
into hard and soft processes is arbitrary and depends on the factorization prescription.
For example, as one increases $M$ a part of the contribution from $g\gamma^* \to q\overline{q}$ where
the anti-quark is nearly collinear to the proton direction is 
reinterpreted as a $q\gamma^* \to q$ process with a modified $f_q(\xi)$, obtained by solving the DGLAP
evolution equations.  Of course in the
case of a $2+1$-jet cross section with $y_{cut}$ sufficiently large that widely separated jets are obtained
these ambiguities are lessened, because this implies a large separation
between the products of the hard and soft parts of the process.

The reason it becomes interesting here to consider this two-scale PMS procedure is that,
for $M$ of order the hard scale $Q$, the quark
and gluon channels give quite different contributions to the weighted integrals in \refeq{meany},
to the extent that the NLO contributions are often of different signs.  There can be large cancellations between them, and it seems plausible that
we could get improved results by ``optimising'' them separately (because the cancellations may well be specific to NLO).

Therefore, we will consider two variants of the PMS in this paper: \PMSa{} where we optimise with respect to a single renormalization
scale $\mu$ and the factorization scale $M$, and \PMSb{} where we optimise with respect to renormalization scales $\mu_q$ and $\mu_g$ for the
quarks and gluons respectively, along with $M$.  It is also possible to assign different scales to the different flavours of quarks,
but this makes no significant difference to the results because the optimisation procedure always chooses very similar scales for them.
Therefore, for simplicity we confine ourselves to considering $\mu_q$ and $\mu_g$.

For \PMSa{} we can define $\mean{y}$ in the conventional way, by expanding the RHS of \refeq{meany} in $\alpha_{\MSb}(\mu)$ and truncating it at NLO.
However, this won't work for \PMSb{} because there are two different couplings, $\alpha_{\MSb}(\mu_q)$ and $\alpha_{\MSb}(\mu_g)$, appearing in both the
numerator and the denominator.  Instead we can perform a double expansion in $\alpha_{\MSb}(\mu_q)$ and $\alpha_{\MSb}(\mu_g)$, keeping all terms with
two or fewer $\alpha$'s.  (We could also truncate the numerator and denominator to NLO and use the quotient as our NLO approximation; this makes
no qualitative difference to the results of this paper).

In performing the optimisation some very low $\MSb$ scheme scales can be obtained (even down to $<1$GeV),
which raises the question as to whether mass thresholds should be taken into account (as
well as the more general question of the validity of the results, which we hope to assess later
by comparison to the data).
The problem is that doing so conflicts with the idea that $\mu$ is a strictly unphysical parameter - indeed, adding
thresholds at any given quark masses would break the scheme invariance of the PMS results.  Another way
of looking at this is to point out that the optimisation is really being performed with respect to the
dimensionless parameter $\tau = b \ln(\mu/\Lambda)$ \cite{Stevenson:1981vj}; varying
$\mu$ at fixed $\Lambda$ is just a convenient
way to achieve this.  Therefore, in this paper we treat the number of flavours as fixed to 5,
as suggested by the value of the physical scale $Q$.

\section{Calculational Methods}
\label{se:analysis}

NLO perturbative predictions for our observables were obtained from the matrix element integration Monte-Carlo
program DISENT 0.1 \cite{Catani:1996jh}, using the interface library NLOLIB \cite{NLOLIB}.
Except where noted, the MRST2001E PDF set \cite{Martin:2002aw} was used, and
$\alpha_{\MSb}(M_Z)$ was fixed to $0.119$ to be consistent with these PDFs.
To allow for optimisation with respect to $M$, many runs of DISENT were carried out at intervals $\Delta M=1$GeV.  
For every value of $Q$ and $M$, $10^8$ events were generated to ensure the Monte-Carlo integration errors were negligible.
The $\mu$ dependence was implemented by
having DISENT compute the coefficients of each power of $\alpha_{\MSb}$ with $\mu=Q$; the $\mu$ logs
as well as the factors of $\alpha_{\MSb}$ could then be added in later. This means that only one run of DISENT was required for all values of $\mu$.
However, to allow this to work it was necessary for all the events to be generated at the same value of $Q$, so the mean
value of $Q$ was used for each bin rather than integrating over the entire width of the bin.  The effect of this was estimated by comparing
the two results for $\mu=M=Q$, giving a correction factor (generally of order $1\%$) which could be applied to the data.

The power corrections can simply be added to the perturbative prediction for the shape means
\be
\mean{y}=\mean{y}_{\mathrm{P}} + \mean{y}_{\mathrm{PC}}.
\ee
We will present fits both for a simple $C_1/Q$ or $C_2/Q^2$ term, and for the more sophisticated universal formula derived in
Refs. \cite{Dokshitzer:1995zt,Dokshitzer:1997ew,Milan,WebDas}.  According to this model
\begin{eqnarray}
\label{eq:DWPC}
\mean{y}_{\mathrm{PC}} & = &a_y \frac{32}{3\pi^2} \frac{\mathcal{M}}{p} \left( \frac{\mu_I}{Q} \right)^p
\left[ \overline{\alpha}_{p-1}(\mu_I) - \alpha_{\MSb}(\mu) - \frac{\beta_0}{2\pi} \right. \nonumber \\
&& \left. (\log(\mu/\mu_I) + K/\beta_0 + 1/p) \alpha_{\MSb}^2(\mu) \right].
\end{eqnarray}
The overall constant $a_y$ and the integer $p$ depend on the observable.  $\overline{\alpha}_{p-1}(\mu_I)$ is the $(p-1)$th
moment of the coupling for $0 < \mu < \mu_I$.  $\mathcal{M} \simeq 1.49$ is the so-called Milan factor \cite{Milan}, and $K$ is
\be
K=\frac{67}{6}-\frac{\pi^2}{2}-\frac{5}{9}N_f.
\ee
The terms in $\mean{y}_{\mathrm{PC}}$ involving $\alpha_{\MSb}$ are intended to subtract the part of the soft-gluon effects that is already included in the
perturbative calculations.  These are to be evaluated with $N_f=5$ and with $\mu$ the renormalization scale used in the
perturbative part.  Note that for our PMS optimised results this will in general differ from $Q$.  In the case of
\PMSb{} the appropriate scale is $\mu_q$ as the power corrections do not receive contributions from the gluon-initiated
sub-process \cite{WebDas}.

For all our observables apart from $\ykt$, $p=1$ and $a_F$ has been calculated, so we can use this formula to perform
a fit for $\overline{\alpha}_0$.  For $\ykt$, $p=2$ and $a_F$ is unknown, which prevents a reliable extraction of
$\overline{\alpha}_1$.  The coefficients are:
\be
a_{\tauc}=1 \;\;\;\; a_{\CP}=\frac{3\pi}{2} \;\;\;\; a_{\rhoE}=\frac{1}{2} \;\;\;\; \cite{WebDas} \nonumber
\ee
\be
a_{\yfJ}=1 \;\; \cite{yfJ} \;\;\;\; a_{\BP}=\frac{\pi}{4\sqrt{\frac{8}{3}\alpha_{\mathrm{CMW}(e^{-3/4}Q)}}} +\frac{3}{8} - \frac{\beta_0}{32}-0.3069+\mathrm{O}(1) \;\; \cite{BP}
\ee

In performing these fits we need to consider both experimental and theoretical sources of error.
The experimental errors consist of both a statistical and a systematic component; lacking knowledge of the
proper correlation between the systematic errors we have treated them as uncorrelated, and simply
added them to the statistical errors in quadrature, to arrive at a composite experimental error.  Where the errors are asymmetric, the
maximum value was taken.

The MRST2001E PDFs allow a PDF-related error to 
be estimated by sampling from an ensemble of different distributions.  Ideally we would repeat the entire analysis
with the different PDFs, but this is impractical because of needing to rerun DISENT for each value of $M$.
Instead we have estimated the error by comparing the analyses at $M=Q$ only.  However, the results are
so stable with respect to changes of $M$ that this should not be too much of a restriction (and in any case,
the errors due to uncertainties on the PDFs are small).

\section{Case Study: $\tauc$}
\label{se:CaseStudy}

In this section we show the results of our analysis, along the lines described in the preceding sections, when applied
to one typical observable, $1-T_c=\tauc$.  A summary of the results for all the observables is given in the next section.

In order to arrive at our \PMSa{} predictions, we need to examine the dependence of $\mean{\tauc}$ on $\mu$ and $M$.
This is illustrated in Fig.~\ref{fg:contours}.  Note that the PMS points, defined by \refeq{PMS}, are saddle points.  Specifically,
they are maxima in the $\mu$-direction and minima in the $M$-direction.  The actual PMS scales are given in Table 1.

The effect of choosing these scales over the standard choice $\mu=M=Q$ is illustrated in Fig.~\ref{fg:taucplot}.
Evidently a substantial power correction is required to fit the data even when using the PMS scales.

\FIGURE{
\includegraphics[scale=0.7]{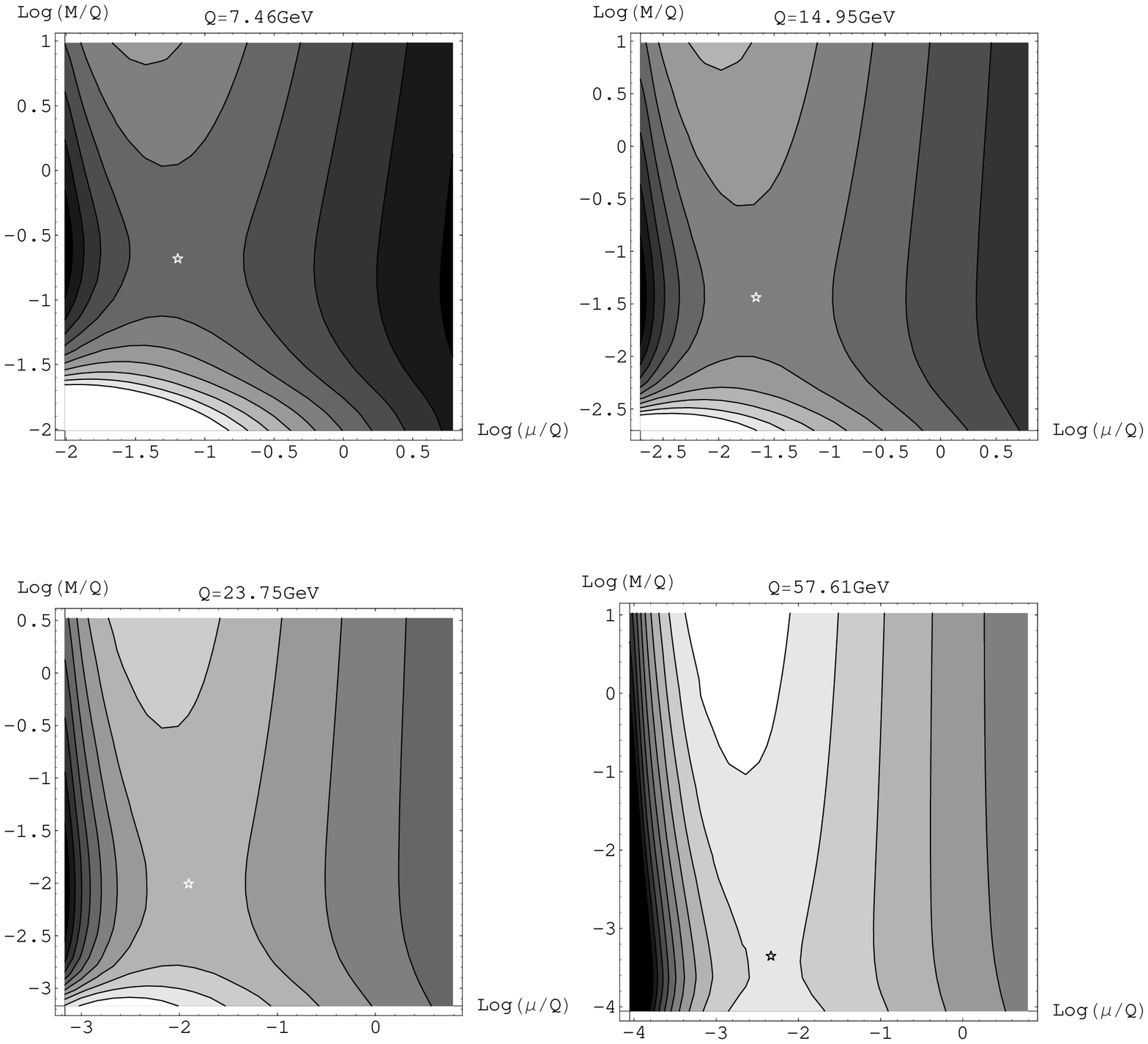}
\caption{\label{fg:contours} Dependence of $\mean{\tauc}^{\mathrm{(NLO)}}$ on $\mu$ and $M$ for various values of $Q$.  In each case, the PMS point is marked.}
}

\TABLE{
\begin{tabular}{|l|c|c|c|c|c|c|c|c|}
\hline
$Q$/GeV & 7.46 & 8.8 & 14.95 & 17.73 & 23.75 & 36.69 & 57.61 & 80.76 \\
\hline
$\mu$/GeV & 2.11 & 2.45 & 4.07 & 4.96 & 6.98 & 12.3 & 19.8 & 22.4 \\
\hline
$M$/GeV & 7.87 & 7.01 & 6.69 & 6.25 & 5.55 & 3.82 & 2.72 & 2.35 \\
\hline
\end{tabular}
\caption{\PMSa{} scales for $\mean{\tauc}$.}
}

\FIGURE{
\includegraphics[scale=1]{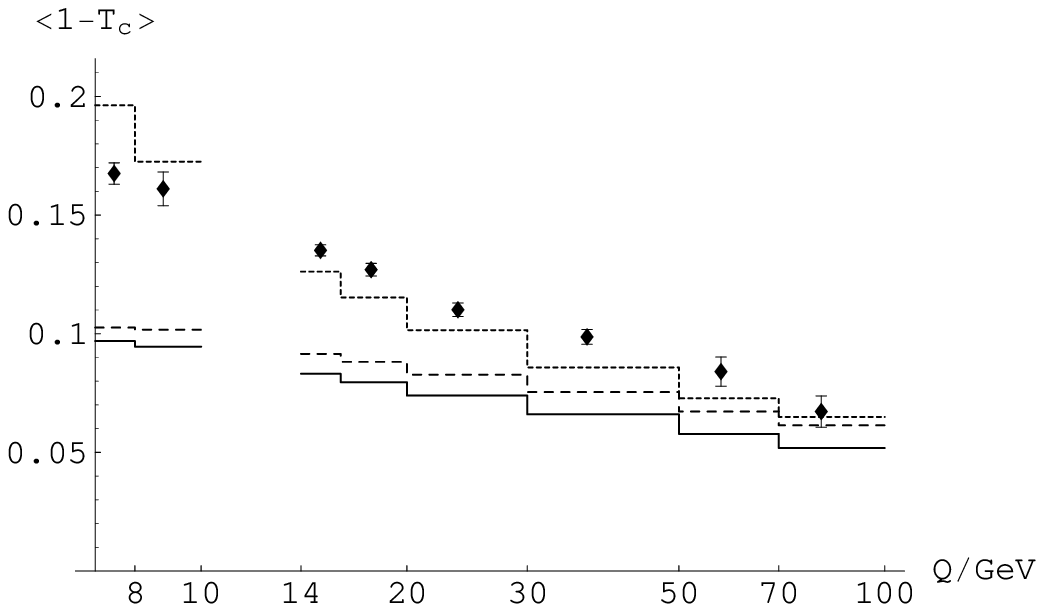}
\caption{\label{fg:taucplot} Comparison of NLO predictions to H1 data \cite{Herameans} for $\mean{\tauc}$
(combined statistical and systematic error bars are shown).
The solid curve uses $\mu=M=Q$.  The dashed curve uses the \PMSa{} scale
choices given in Table 1.  The dotted curve uses the \PMSb{} scale choices shown in Table 2.}
}

To evaluate the \PMSb{} approximation requires one to look at the dependence of $\mean{\tauc}$ on $\mu_q$, $\mu_g$ and
$M$.  Some typical examples are illustrated in Fig.~\ref{fg:PMSbillustration}.
The PMS points here are maxima in the $\mu_q$ and $\mu_g$ directions, and may be either minima or maxima in the $M$ direction;
the relevant scales are listed in Table 2.  There is a clear difference between the values of $\mu_q$ and $\mu_g$, the former are very small
and the latter very large.
The result of using these \PMSb{} scales is shown as the dotted
line on Fig.~\ref{fg:taucplot}.  Clearly, at least for large $Q$, the size of the required power correction is
substantially reduced, but at low $Q$ the prediction now actually {\it overshoots} the measurement.  It is perhaps
not surprising that we run into trouble here, as $\mu_q$ is extremely low, in fact around 0.5GeV$\simeq 2\Lambda_{\MSb}$, so $\alpha_{\MSb}(\mu_q) \simeq 0.7$.

To see why \PMSb{} makes such a large difference compared to \PMSa{} it is helpful to fix $M=Q$ - (it turns
out that the effect of optimising $M$ is rather small, and the relationship between \PMSa{} and \PMSb{}
is simplest when they have a common $M$).
Consider the coefficients in the perturbative expansion for the integral
{\setlength\arraycolsep{1pt}
\begin{eqnarray}
\label{eq:WI}
\int dX \tauc(X) \frac{d\sigma(ep\to X)}{dX}=&&A_q \alpha_s(\mu_q) + \left(B_q + \beta_0 \log\left(\frac{\mu_q^2}{Q^2}\right) A_q\right) \alpha_s^2(\mu_q) + \nonumber \\
&& A_g \alpha_s(\mu_g) + \left(B_g + \beta_0 \log\left(\frac{\mu_g^2}{Q}\right) A_g\right) \alpha_s^2(\mu_g) .
\end{eqnarray}
}%
The coefficients $A$ and $B$ depend on both $M$ and $Q$.  Taking $Q=7.46$GeV where
the difference between \PMSa{} and \PMSb{} is most obvious (and fixing $M=Q$ as noted above) we have
\be
\label{eq:cancellations}
A_q = 5.29,\, A_g = 1.44, \;\;  B_q = 15.4, \, B_g = -10.3.
\ee
On the other hand, if $\mu_q$ and $\mu_g$ are identified as in \PMSa{}, the coefficients in the perturbation series are simply
\be
A = A_q + A_g = 6.74 \;\;\;\; B=B_q+B_g = 5.10.
\ee
Because of the cancellation in the NLO coefficient $B$, \PMSa{} sees a series which appears to be much more convergent 
than that seen by \PMSb{}, and this lessens the effect of the optimisation (this can be understood most easily by recalling the similarity
between PMS and the Method of Effective Charges, which fixes the scale so that $B=0$).  
Therefore, it may be that \PMSa{} underestimates the size of higher orders in the $\mu=M=Q$ series.  If this cancellation does not persist
to higher orders, then one would expect \PMSb{} to give a more realistic estimate of the higher order terms.

This explanation for the values of the \PMSb{} scales is an over-simplification, because we aren't actually optimising the weighted integral \refeq{WI}, but
rather the ratio \refeq{meany}.  However, the total cross-section is convergent enough that these simple considerations
do capture the essential reason behind the PMS scales shown in Tables 1 and 2.  For example, the Effective Charge
scales $\mu_\mathrm{EC}=\exp(-B/(\beta_0A))Q$ corresponding to these coefficients are
{\setlength\arraycolsep{1pt}
\begin{eqnarray}
\mu &=& 4.0\mathrm{GeV}\\
\mu_q = 0.68\mathrm{GeV} && \mu_g = 2.6\cdot10^3\mathrm{GeV}
\end{eqnarray}
}%
which are all rather close to the corresponding PMS scales (although the agreement here is perhaps
deceptively good and worsens at higher energies, the EC $\mu_g$ actually falling as $Q$ is increased).

\FIGURE{
\includegraphics[scale=0.45]{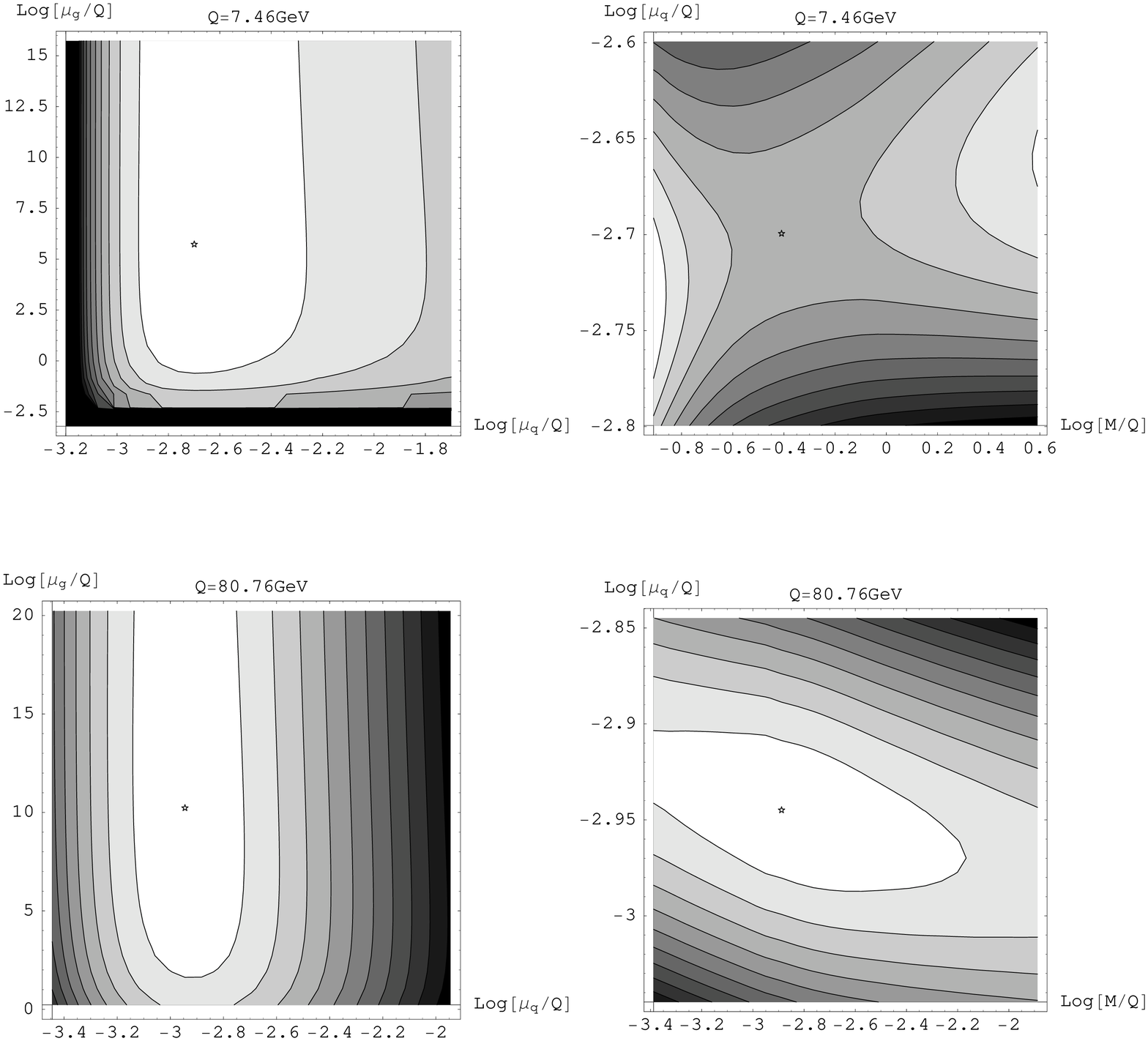}
\caption{\label{fg:PMSbillustration} Dependence of $\mean{\tauc}^{\mathrm{(NLO)}}$ on $\mu_q$, $\mu_g$ and $M$ for the extreme $Q$ values.  
The left hand plot shows the dependence on $\mu_q$ and $\mu_g$ with $M$ fixed to its PMS value.  The right
hand plot shows the dependence on $M$ and $\mu_q$ with $\mu_g$ fixed to its
PMS value.  The PMS points are marked, and the numerical values of the PMS scales are listed in Table 2.
}
}

\TABLE{
\begin{tabular}{|l|c|c|c|c|c|c|c|c|}
\hline
$Q$/GeV & 7.46 & 8.8 & 14.95 & 17.73 & 23.75 & 36.69 & 57.61 & 80.76 \\
\hline
$\mu_q$/GeV & 0.50 & 0.59 & 0.90 & 1.1 & 1.4 & 2.0 & 3.2 & 4.2 \\
\hline
$\mu_g$/GeV & 2.3$\cdot10^3$ & 6.0$\cdot10^3$ & 6.5$\cdot10^4$ & 1.5$\cdot10^5$ & 4.1$\cdot10^5$ & 1.0$\cdot10^6$ & 1.9$\cdot10^6$ & 2.2$\cdot10^6$ \\
\hline
$M$/GeV & 4.95 & 5.20 & 6.23 & 6.41 & 6.76 & 6.70 & 5.79 & 4.50 \\
\hline
\end{tabular}
\caption{\PMSb{} scales for $\mean{\tauc}$.}
}

$\tauc$ is expected to receive $1/Q$ power corrections, which we can try to describe either
by simply adding a term $C_1/Q$ to the perturbative predictions, or by using \refeq{DWPC} which
relates the corrections to $\alphazero$.  Because of the rise in the \PMSb{} predictions
for $Q<10$GeV, these $1/Q$ corrections alone cannot compensate for the discrepancy between theory and data in this
case.  In addition, there must be large higher-order $\mathrm{O}(\alpha_s^3)$ effects and/or sub-leading
power corrections $\propto 1/Q^2$.  To compare the size of the power corrections required by the different
perturbative predictions, it makes sense to exclude low $Q$ data if this gives an unacceptably bad fit.
This is especially true in view of the fact that the $e^+e^-$ data examined in Ref.~\cite{r12} had $\mean{Q}>45$GeV,
so these additional effects might be important at low $Q$ in the $e^+e^-$ case also.
Therefore, we performed minimum-$\chi^2$ fits, adding in data points from the highest $Q$ downwards until the
fit probability fell below 5\%.  Experimental errors were estimated by allowing $\chi^2$ to vary within $1$ of its minimum value.
Fitting in this way for $C_1$ gives
{\setlength\arraycolsep{1pt}
\begin{eqnarray*}
\mu=M=Q:\;\; &C_1=1.23(11)(20)\GeV,&\;\;Q>30\GeV \\
\mathrm{PMS}_1:\;\; &C_1=0.65(2)(2)\GeV,&\;\;Q>8\GeV \\
\mathrm{PMS}_2:\;\; &C_1=0.18(3)(5)\GeV,&\;\;Q>14\GeV.
\end{eqnarray*}}%
Here and throughout, the first number in brackets indicates the error due to experimental and PDF uncertainties;
the second number gives an indication of the sensitivity to the fit range by showing the
size of the shift induced by excluding the lowermost bin.
As expected, the required power correction is largest for the ``physical scale'', and smallest for \PMSb{}.
Of the three predictions, \PMSa{} gives the best description of the data, fitting the largest range of data
with the best stability.

Because one expects also sub-leading power corrections  to be present, it is interesting
to introduce e.g. a $C_2/Q^2$ term into the fit, to see how this affects the conclusions:
{\setlength\arraycolsep{1pt}
\begin{eqnarray*}
\mu=M=Q:\;\; &C_1=1.09(5)(6)\GeV,\;C_2=-4.3(5)(9)\GeV^2, &\;\;Q>7\GeV \\
\mathrm{PMS}_1:\;\; &C_1=0.82(5)(1)\GeV,\;C_2=-2.5(5)(2)\GeV^2, &\;\;Q>7\GeV \\
\mathrm{PMS}_2:\;\; &C_1=0.49(5)(0)\GeV,\;C_2=-5.3(5)(1)\GeV^2, &\;\;Q>7\GeV
\end{eqnarray*}}%
where the errors are strongly correlated.  Unsurprisingly, the $C_2/Q^2$ term
allows even the low $Q$ data to be correctly described.  The basic fact that the
optimisation reduces the need for $1/Q$ power corrections does seem to survive.

Fitting for $\alphazero$ gives
{\setlength\arraycolsep{1pt}
\begin{eqnarray*}
\mu=M=Q:\;\; &\alphazero=0.524(7)(5),&\;\;Q>8\GeV \\
\mathrm{PMS}_1:\;\; &\alphazero=0.596(6)(7),&\;\;Q>7\GeV \\
\mathrm{PMS}_2:\;\; &\alphazero=0.614(8)(16),&\;\;Q>14\GeV.
\end{eqnarray*}}%
It may perhaps seem surprising that the $\alphazero$ values are larger for the PMS scale choices,
even though the size of the power corrections seems to be reduced, as is reflected in the $C_1$ values.
This is a consequence of using the PMS scales in \refeq{DWPC}, which increases the perturbative contribution
and requires larger $\alphazero$ to compensate.

Finally, we can examine the effect of assuming a different value for $\alpha_{\MSb}$.  For consistency, this requires
the PDFs to be changed.  Fig.~\ref{fg:vary_alpha} shows the effects on $\mean{\tauc}$ of choosing $\alpha_{\MSb}(M_Z)=0.1125$ and
$0.1225$, using PDFs from Ref.~\cite{MRST_vary}.  To save computer time $M$ was fixed to $Q$ throughout.
This makes little difference to the results - 
indeed, this can now easily be seen by comparing the central graph of Fig.~\ref{fg:vary_alpha} with  Fig.~\ref{fg:taucplot}, which only differ in
that $M$ is fixed to $Q$ in the former and optimised in the latter.

The results of performing power corrections fits for these values of $\alpha_{\MSb}$ are summarised in Table 3.
The effect of the optimisation is still to reduce the size of the power corrections (although less substantially
for low $\alpha_{\MSb}$ as would be expected).

\FIGURE{
\parbox{100mm}{\vbox{\vskip-1cm\includegraphics[scale=0.8]{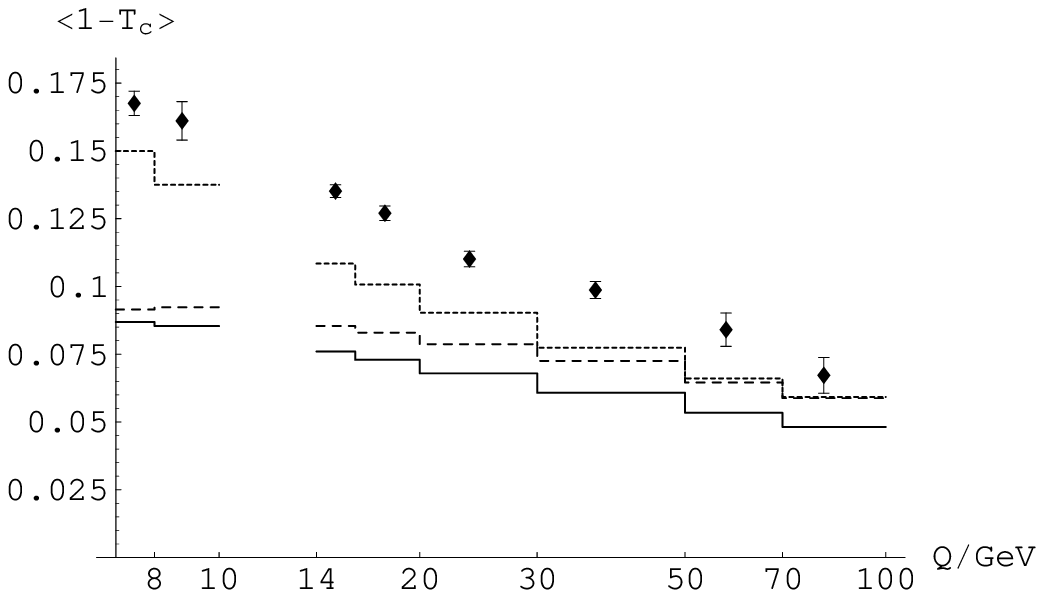}\vskip-3cm\includegraphics[scale=0.8]{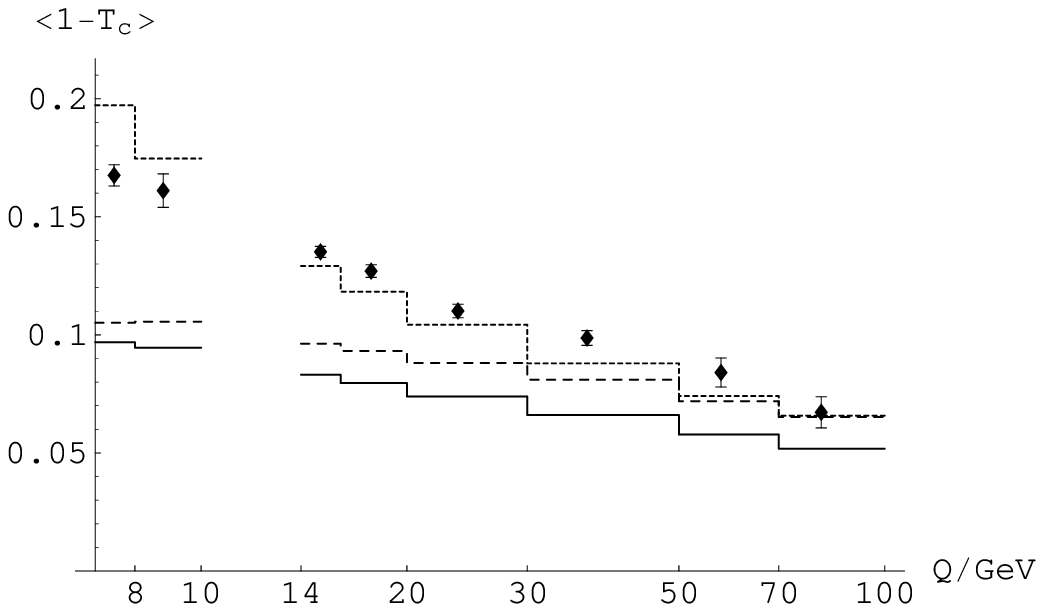}\vskip-3cm\includegraphics[scale=0.8]{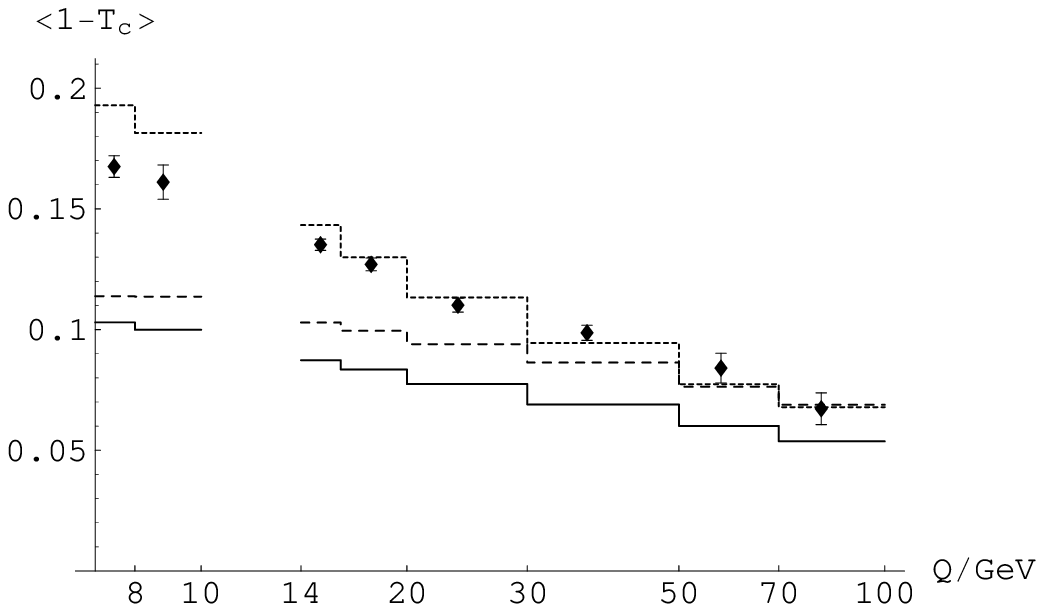}\vskip-1cm}}
\caption{\label{fg:vary_alpha} Effect of varying $\alpha_{\MSb}$ on the values of $\mean{\tauc}$.
The top graph has $\alpha_{\MSb}(M_Z)=0.1125$, the middle has $\alpha_{\MSb}(M_Z)=0.119$, and the bottom has $\alpha_{\MSb}(M_Z)=0.1225$.
Throughout we have fixed $M=Q$.  The solid line has $\mu=Q$, the dashed line has $\mu$ picked by \PMSa{}, and the dotted line has $\mu$ picked by \PMSb{}.  Comparing the central graph with Fig.~\ref{fg:taucplot} shows the effect of
optimising $M$.}
}

\TABLE{
\begin{tabular}{|c|c|c|c|c|c|c|}
\hline
\multicolumn{7}{|c|}{$C_1/$GeV} \\
\hline
$\alpha_{\MSb}(M_Z)$ & \multicolumn{2}{|c|}{$\mu=Q$} & \multicolumn{2}{|c|}{\PMSa{}} & \multicolumn{2}{|c|}{\PMSb{}}\\
\hline
$0.1125$ & $1.43(11)(27)$ & $>30$ & $0.74(2)(2)$ & $>8$ & $0.51(4)(6)$ & $>16$ \\ 
 \hline 
$0.119$ &  $1.23(11)(20)$ & $>30$ & $0.53(2)(3)$ & $>7$ & $0.13(3)(5)$ & $>14$ \\ 
 \hline 
$0.1225$ & $0.87(6)(24)$ & $>20$ & $0.44(2)(2)$ & $>7$ & $-0.09(2)(1)$ & $>8$ \\ 
 \hline 
\multicolumn{7}{|c|}{$\alphazero$} \\
\hline
$0.1125$ & $0.661(33)(69)$ & $>30$ & $0.586(7)(5)$ & $>8$ & $0.559(6)(4)$ & $>7$ \\ 
 \hline 
$0.119$ & $0.524(7)(5)$ & $>8$ & $0.596(6)(6)$ & $>7$ & $0.575(8)(15)$ & $>14$ \\ 
 \hline 
$0.1225$ & $0.511(6)(9)$ & $>7$ & $0.615(6)(2)$ & $>7$ & $0.519(18)(5)$ & $>20$ \\ 
 \hline 
\end{tabular}
\caption{Fits to $\mean{\tauc}$ for $C_1$ and $\alphazero$, with various value of $\alpha_{\MSb}(M_Z)$.
Note that $M=Q$ throughout.
The first number gives the best fit value
and numbers in brackets indicate errors in the last digits (due to
experimental and PDF uncertainties) and the variation arising from
removing one bin from the fit.  The notation $>Q$ indicates the
range of $Q$ that could be fitted before the $\chi^2$ indicated a fit probability
of $<5\%$.}
}

\section{Results}
\label{se:Results}

In this section we summarise results for all the observables studied in Ref.~\cite{Herameans}.
The perturbative predictions are compared to data in Figs.~\ref{fg:results1} and \ref{fg:results2}.
$\tauc$ and $\CP$ show similar features: the \PMSa{} predictions are somewhat closer to the data
than the physical scale ones, and the \PMSb{} predictions are a lot closer, except they become too
large at low energies.  The PMS predictions for $\tauP$ are closer to the data than those with the
$\mu=M=Q$,
with no evident breakdown at low $Q$.  $\BP$ and $\rhoE$ also have reasonable low $Q$ behaviour
but with a lesser improvement of the fit.  Turning to $\rhoEE$, though, the improvement
(especially for \PMSb{}) is much more substantial.
The jet transition parameters, $\yfJ$ and $\ykt$, move {\it away} from the data when the scales are optimised.

\FIGURE{
\includegraphics[scale=0.8]{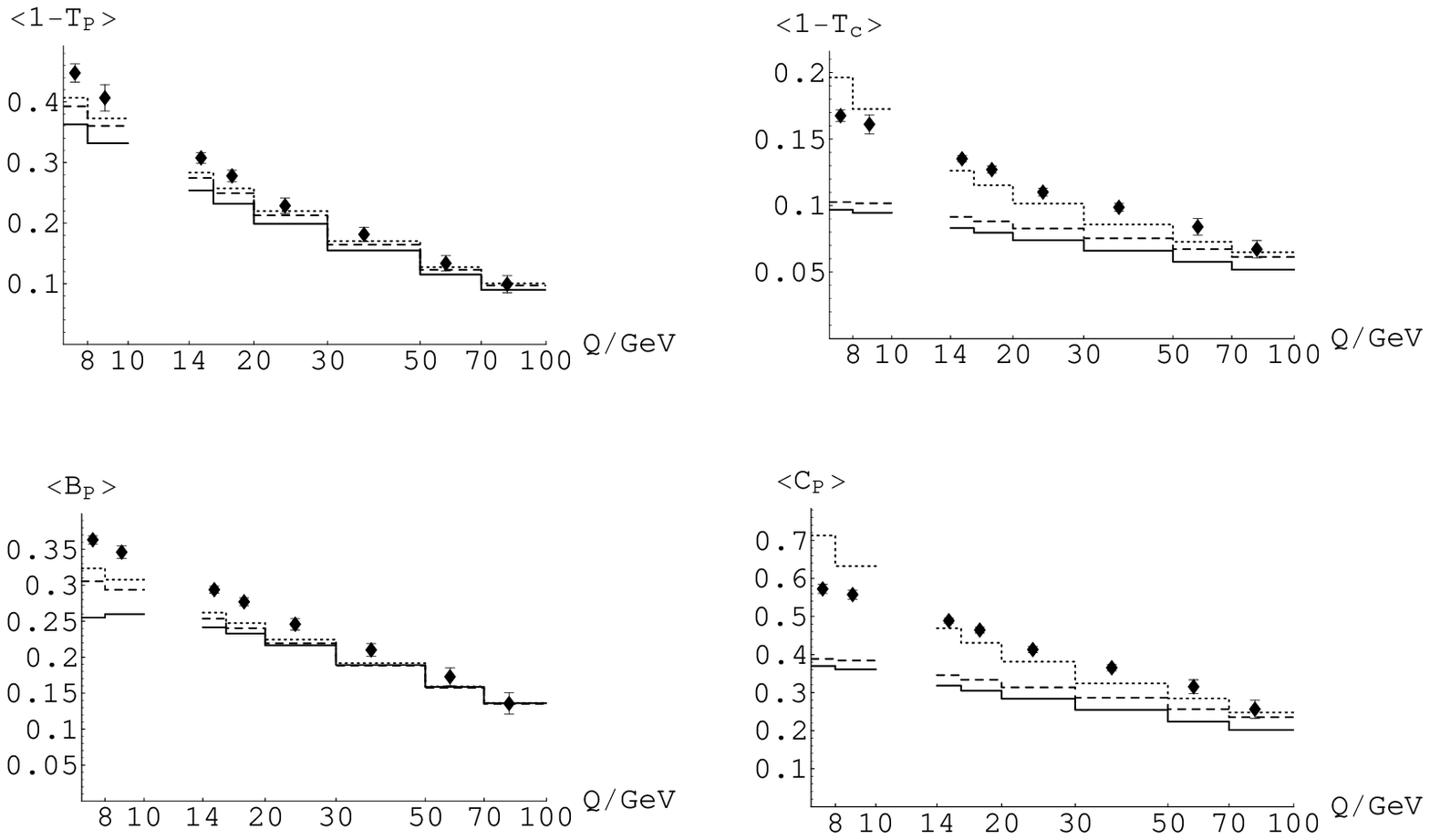}
\caption{\label{fg:results1} Predictions of our three perturbative predictions compared to data.
The solid line uses the scale choice $\mu=M=Q$, the dashed curve uses \PMSa{} and the dotted curve
\PMSb{}.  Data are shown from Ref.~\cite{Herameans}.
}
}

\FIGURE{
\includegraphics[scale=0.8]{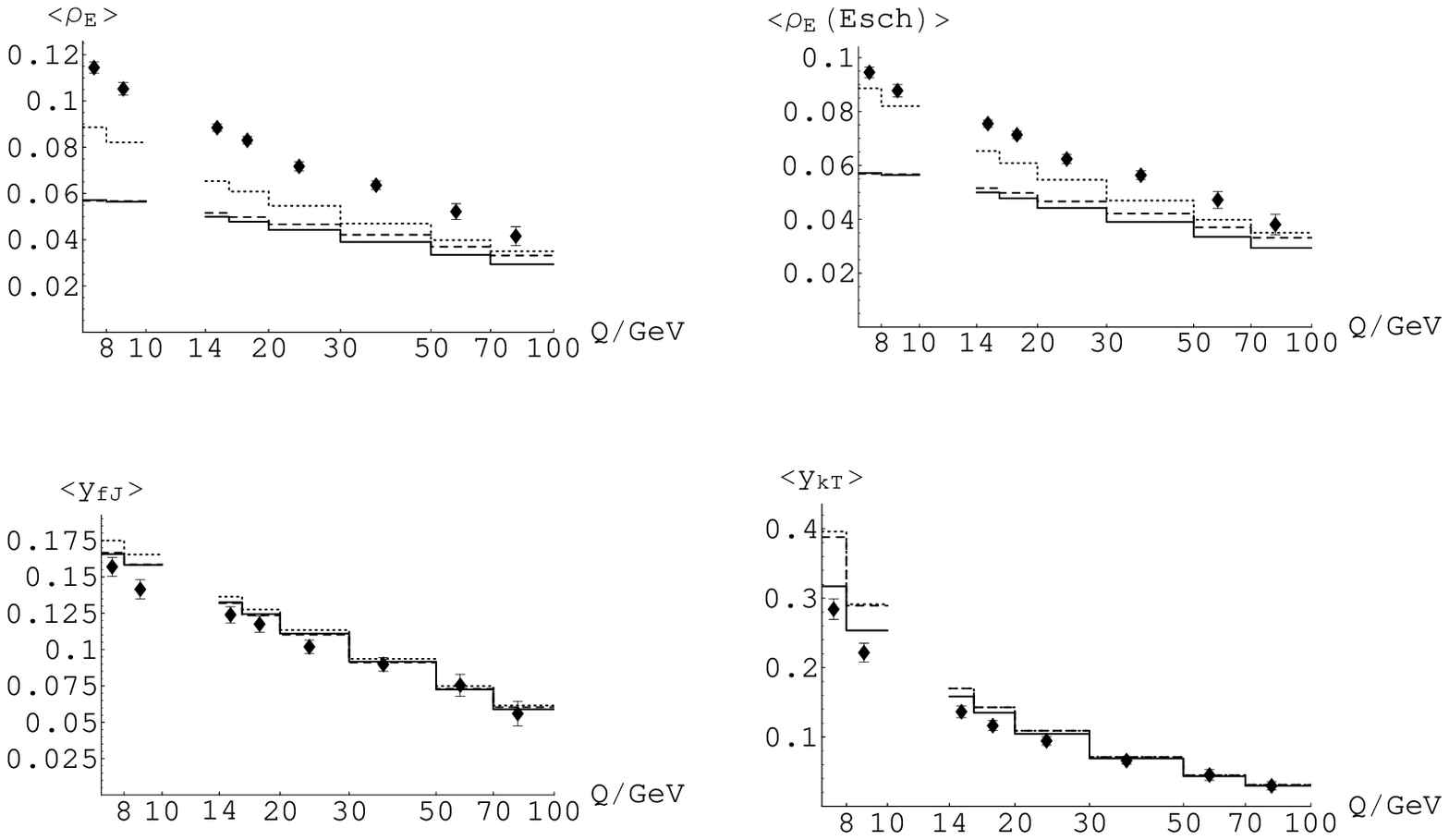}
\caption{\label{fg:results2} Predictions of our three perturbative predictions compared to data.
The solid line uses the scale choice $\mu=M=Q$, the dashed curve uses \PMSa{} and the dotted curve
\PMSb{}.  Data are shown from Ref.~\cite{Herameans}.
}
}

Table 4 shows fits for a $C_1/Q$ power correction for all
observables except $\ykt$ and a fit for a $C_2/Q^2$ power correction for $\ykt$
($\ykt$ is the only observable whose leading power correction is expected to be $\propto 1/Q^2$).
Table 5 shows the results of fitting for a power correction based on \refeq{DWPC} for all observables bar $\ykt$.
These fits allow us to see to what extent the discrepancy between the perturbative predictions and the data visible
in Figs.~\ref{fg:results1} and \ref{fg:results2} can actually be described as a power correction.
As was noted in Section \ref{se:CaseStudy}, \PMSb{} fails to give a good description of $\tauc$ at low $Q$.
This is also the case for $\CP$ and $\rhoE$ (and $\rhoEE$).  
Although not obvious from Figs.~\ref{fg:results1} and \ref{fg:results2}, the ``physical scale'' predictions 
don't describe the low energy data for these observables very well either.  In fact, the best overall fits seem to be those
that use \PMSa{}.
The other observable with a large discrepancy at low $Q$, $\ykt$, can
be described quite well by any of the scale choices provided we add the expected $1/Q^2$ power correction.

\TABLE{
\begin{tabular}{|c|c|c|c|c|c|c|}
\hline
\multicolumn{7}{|c|}{$C_1/$GeV} \\
\hline
Obs & \multicolumn{2}{|c|}{$\mu=M=Q$} & \multicolumn{2}{|c|}{\PMSa{}} & \multicolumn{2}{|c|}{\PMSb{}}\\
\hline
$\tauP$ & $0.73(7)(5)$ & $>7$ & $0.45(7)(2)$ & $>7$ & $0.33(7)(1)$ & $>7$ \\ 
 \hline 
$\tauc$ & $1.23(11)(20)$ & $>30$ & $0.65(2)(2)$ & $>8$ & $0.18(3)(5)$ & $>14$ \\ 
 \hline 
$\BP$ & $0.79(3)(2)$ & $>7$ & $0.50(3)(7)$ & $>7$ & $0.37(3)(8)$ & $>7$ \\ 
 \hline 
$\CP$ & $4.17(29)(92)$ & $>30$ & $2.26(7)(15)$ & $>14$ & $0.76(49)(24)$ & $>16$ \\ 
 \hline 
$\rhoE$ & $0.92(6)(13)$ & $>30$ & $0.67(4)(12)$ & $>20$ & $0.49(4)(13)$ & $>20$ \\ 
 \hline 
$\yfJ$ & $-0.11(3)(3)$ & $>7$ & $-0.11(3)(2)$ & $>7$ & $-0.18(3)(3)$ & $>7$ \\ 
 \hline 
$\rhoEE$ & $0.65(5)(12)$ & $>30$ & $0.38(2)(2)$ & $>14$ & $0.21(2)(4)$ & $>16$ \\ 
 \hline 
\multicolumn{7}{|c|}{$C_2/$GeV$^2$} \\
\hline
 $\ykt$ & $-2.70(60)(87)$ & $>7$ & $-6.03(60)(24)$ & $>7$ & $-6.30(60)(6)$ & $>7$ \\ 
\hline 
\end{tabular}
\caption{Fits for $C_1$ and $C_2$ to all observables. See Table 3 for an
explanation.}
}

\TABLE{
\begin{tabular}{|c|c|c|c|c|c|c|}
\hline
\multicolumn{7}{|c|}{$\alphazero$} \\
\hline
Obs & \multicolumn{2}{|c|}{$\mu=M=Q$} & \multicolumn{2}{|c|}{\PMSa{}} & \multicolumn{2}{|c|}{\PMSb{}}\\
\hline
$\tauP$ & $0.519(22)(2)$ & $>7$ & $0.533(22)(20)$ & $>7$ & $0.559(22)(27)$ & $>7$ \\ 
 \hline 
$\tauc$ & $0.524(7)(5)$ & $>8$ & $0.596(6)(7)$ & $>7$ & $0.614(8)(16)$ & $>14$ \\ 
 \hline 
$\BP$ & $0.568(17)(20)$ & $>8$ & $0.447(13)(9)$ & $>7$ & $0.443(13)(6)$ & $>7$ \\ 
 \hline 
$\CP$ & $0.465(7)(12)$ & $>16$ & $0.527(3)(6)$ & $>7$ & $0.557(7)(23)$ & $>16$ \\ 
 \hline 
$\rhoE$ & $0.808(38)(66)$ & $>30$ & $0.715(10)(15)$ & $>14$ & $0.736(7)(9)$ & $>7$ \\ 
 \hline 
$\yfJ$ & $0.264(18)(22)$ & $>7$ & $0.261(10)(17)$ & $>7$ & $0.278(12)(19)$ & $>8$ \\ 
 \hline 
$\rhoEE$ & $0.642(34)(56)$ & $>30$ & $0.572(7)(15)$ & $>8$ & $0.611(7)(10)$ & $>8$ \\ 
 \hline
\end{tabular}
\caption{Fits for $\alphazero$.  See Table 3 for an explanation.}
}

\section{Conclusions}
\label{se:Conclusions}

In this paper we have studied how power correction fits to event shape means in DIS are affected by
choosing the factorization and renormalization scales according to the Principle of Minimal Sensitivity.
In doing this, two different prescriptions were adopted: \PMSa{}, where the unphysical parameters
were taken to be $\mu$ and $M$ and \PMSb{}, where different values of $\mu$ were used for
the quark- and gluon-initiated sub-processes.  The motivation behind \PMSb{} was to
avoid underestimating the effect of higher order corrections because of the cancellations
between the $q\gamma^*$ and $g\gamma^*$ sub-processes at NLO (illustrated in \refeq{cancellations}).

\PMSa{} gives results that are pretty close to those found using the conventional choice $\mu=M=Q$.
However, it does improve the quality of the power correction fits (see Tables 4 and 5).
\PMSb{} gives perturbative results that are substantially closer to the data at high values of $Q$, but
which deviate from it at low $Q$ (see Figs.~\ref{fg:results1} and \ref{fg:results2}).
If we exclude this low $Q$ region
from the fits (as in Tables 4 and 5), \PMSb{} requires smaller power corrections to fit the data than does either
\PMSa{} or the choice $\mu=M=Q$.

In carrying out these calculations we have used the MRST2001E PDF set.
It is possible that problems could arise from using optimisation with PDFs that were obtained without optimisation.  If, say, switching to the
PMS scale has a consistent effect on a number of observables, refitting for the PDFs using the PMS would cause them to change
so as to counteract the effect of the optimisation.  Unfortunately, optimising enough observables to be able to fit for the PDFs would 
be a very large undertaking, and is beyond the
scope of this paper.  One might hope that this is not actually an issue, because event shapes have particularly large NLO corrections
and hence are probably more sensitive to optimisation than most other observables.

The motivation for this study was to determine whether ``optimised'' scales could significantly reduce the
need for power corrections to DIS event shape means as they appeared to do for their $e^+e^-$ counterparts \cite{r12}.
This would suggest that what appears at NLO to be a power correction is just the combined effects of
higher order perturbative terms (although why genuine non-perturbative corrections would be so small is
quite mysterious).
\PMSa{} does not support this idea; \PMSb{} does to some extent, but is poorly behaved at low
energies - however, these energies are much lower
than those studied in \cite{r12}.  So if \PMSb{} provides a better estimate
of higher order terms in the perturbation series than \PMSa{}, it could be that the conclusions of Ref.~\cite{r12} do
extend to DIS event shape means.  In this case, NNLO effects and/or power corrections
would become important in both processes at $Q\simeq 20$GeV.

To try to determine which optimisation is most accurate one can turn to the data.
Unfortunately, fitting for a power correction largely absorbs the differences between them, leaving
only the fact that \PMSa{} gives a better description of the low energy data due to it lacking
the rapid growth of \PMSb{} (visible for example on Fig.\ref{fg:taucplot}).
Another way to test the different optimisations 
would be to perform a PMS analysis of the ZEUS data, some of which is binned in both
$Q$ and $x_B$ \cite{Chekanov:2002xk}.
However, as long as we only have NLO calculations to work with it will be difficult to be sure which,
if any, optimisations are
best (although the overall consistency of the Method of Effective Charges analysis in Ref.~\cite{r12} is highly
suggestive).  Once NNLO computations become available for these event shape means it will be possible to measure the
convergence of the optimised approximants and compare them to each other and to the convergence in the $\MSb$-scheme.
This should provide better guidance in choosing the scheme for these observables, and allow us to more thoroughly test if
the apparent power corrections really can be mimicked by an optimisation of the scheme.

\acknowledgments

The author would like to thank Chris Maxwell for helpful comments on an earlier version of this paper,
Mike Seymour for helpful conversations and PPARC for provision of a UK Studentship.

\end{document}